# IONIZATION PROFILE MONITORS FOR THE IOTA STORAGE RING AT FERMILAB*

R. Thurman-Keup[†,1], N. Banerjee[1], M. Bressler[1], K. Carlson[1], D. Edstrom[1], B. Fellenz[1], M. Mwaniki[2], H. Piekarz[1], A. Romanov[1], A. Semenov[1], V. Shiltsev[3]
[1]Fermi National Accelerator Laboratory, Batavia, IL, USA
[2]Illinois Institute of Technology, Chicago, IL, USA
[3]Northern Illinois University, DeKalb, IL, USA

*Abstract*

Ionization profile monitors (IPM) are used in many accelerator laboratories around the world for non-destructive profile measurements of high-energy particle beams. They have been used in nearly all the synchrotrons built at Fermilab and are presently used in the Main Injector (MI), Recycler (RR), and Booster. This paper will present the current status of the design of a new pair of IPMs to be installed in the Integrable Optics Test Accelerator (IOTA), an experimental proton storage ring with a beam energy of 2.5 MeV. The design incorporates aspects of several IPMs previously or currently used at FNAL. Similar to the Booster IPM, ions are collected rather than electrons, thus requiring no external magnetic field. To increase the signal, each IPM uses two microchannel plates in the chevron configuration as is done in the MI/RR IPMs, and due to the ring's ultrahigh vacuum conditions, the local vacuum pressure is increased with a calibrated leak, akin to the former Tevatron IPMs, to increase the ionization yield per bunch. Data acquisition includes upgraded preamplifiers and leveraging of the digitizer design of the beam position monitor system already in use at IOTA. In addition to the mechanical and electronic designs, simulation results of the effects of the electric fields and injected gas on the circulating beam will also be presented.

## INTRODUCTION

One of the key pieces of information needed for optimization of the beam in a particle accelerator is the transverse profile. A wide variety of transverse profile instruments exist, including physical wire scanners, synchrotron light imagers, optical transition radiation imagers, and gas ionization/fluorescence profile devices. For synchrotrons, in order to preserve beam quality and reduce diagnostic damage, the choices are reduced to non-destructive. Thus, unless the Lorentz factor, $\gamma$, is large enough to generate synchrotron radiation (*e.g.* the Large Hadron Collider), gas-based profiling is the remaining choice. For the integrable optics test accelerator (IOTA) at Fermilab (Fig. 1) [1], a gas ionization profile monitor (IPM) is being constructed to measure profiles of the 2.5 MeV protons in the IOTA ring [2]. This energy level makes this IPM one of the lowest energy applications yet, being slightly less than the 4.2 MeV/u measurements at LEIR [3,4].

The upcoming experimental beam physics program with protons in IOTA includes a broad range of studies — nonlinear beam dynamics, coherent beam instabilities, space charge effects, bunch compression [5], and others — all requiring reliable, fast diagnostics of the evolution of beam sizes and emittances. IOTA's beam lifetime specifications require an average vacuum pressure below $10^{-9}$ Torr in the ring, which impacts the IPM design through the need for external gas injection as well as increased signal amplification.

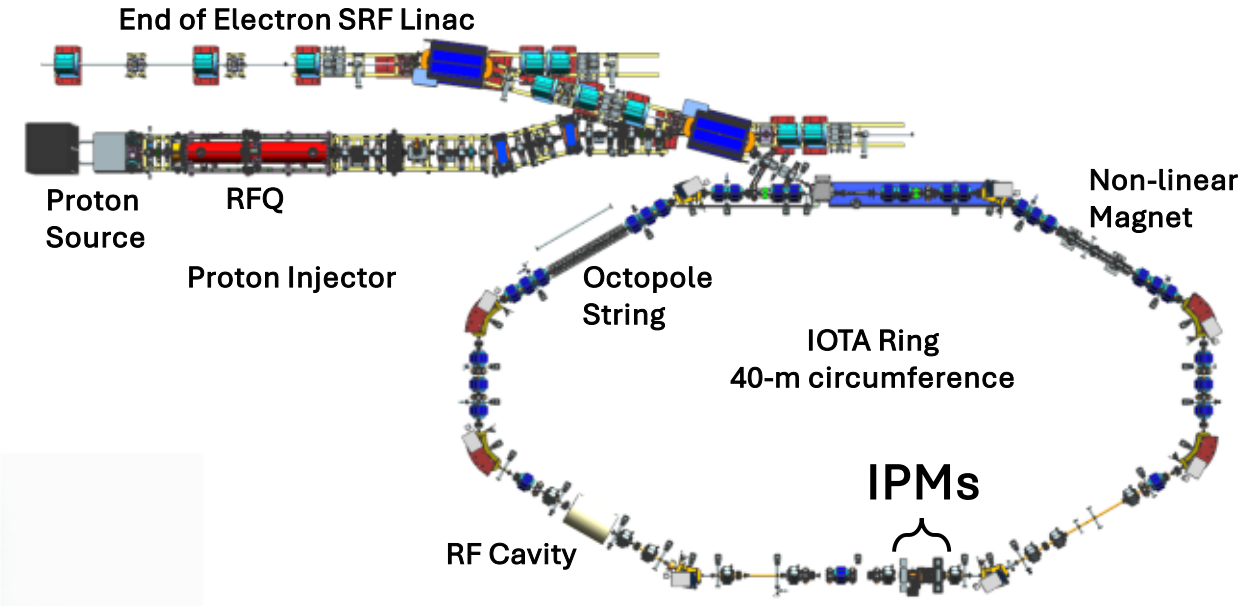

Figure 1: The 40-m-circumference IOTA ring showing injection from the proton injector. The IPMs are located in the bottom straight section of the ring.

Table 1: IOTA Proton Beam Parameters

| Parameter | Design |
|---|---|
| Proton kinetic energy | 2.5 MeV |
| Number of bunches | 1 - 4 |
| Revolution period | 1.83 µs |
| Average current / bunch | 2 mA |
| Transverse rms size @ IPM | 2.5 mm |
| Average vacuum pressure | $6 \times 10^{-10}$ Torr |

## EXPERIMENTAL DEVICE

The IPM design for IOTA is loosely based on the design already in use at the Booster synchrotron and will collect the molecular ions from the gas ionization process (Fig. 2). Relevant parameters are shown in Table 2. The impact on the measured profile due to space-charge-induced spreading of the ions while being collected is estimated to be less than 10% [2]. This fact coupled with the space constraints of the IOTA ring led to the use of ion collection as opposed to electron collection which would require external magnetic fields to constrain the trajectory of the electrons.

† keup@fnal.gov

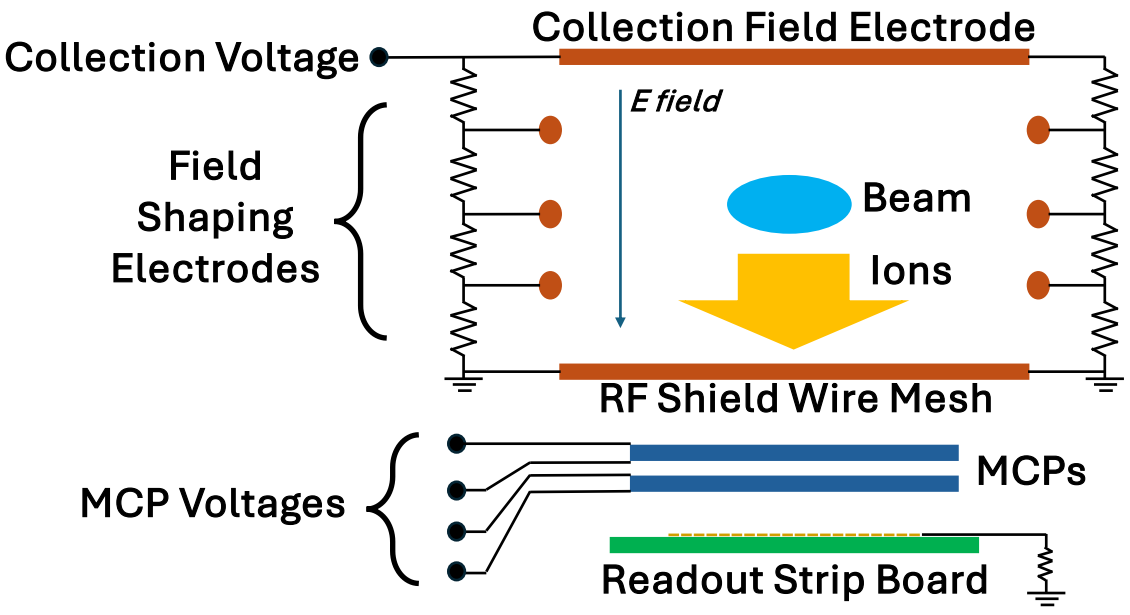

Figure 2: Diagram of the ion collection system. The beam is normal to the page. The ions are directed down through the wire mesh to the MCPs where they generate electrons which are amplified and collected on the strip board.

Table 2: IPM Design Parameters

| Parameter | Design |
|---|---|
| Insertion Length | 0.808 m |
| Collection field gap | 57 mm |
| Signal collection length | 93 mm |
| Readout strips / IPM | 60 |
| Strip width | 0.35 mm |
| Strip center spacing | 0.5 mm |
| Collection voltage | 16 KV |
| IPM-region pressure | $10^{-9}$ Torr |

## *Mechanical Design*

The mechanical structure includes two cylindrical housings that contain the detector components, four Agilent Vacion Plus 55 vacuum pumps at each end, and a dipole corrector magnet between the two cylinders to correct for the electric fields that collect the ions (Fig. 3).

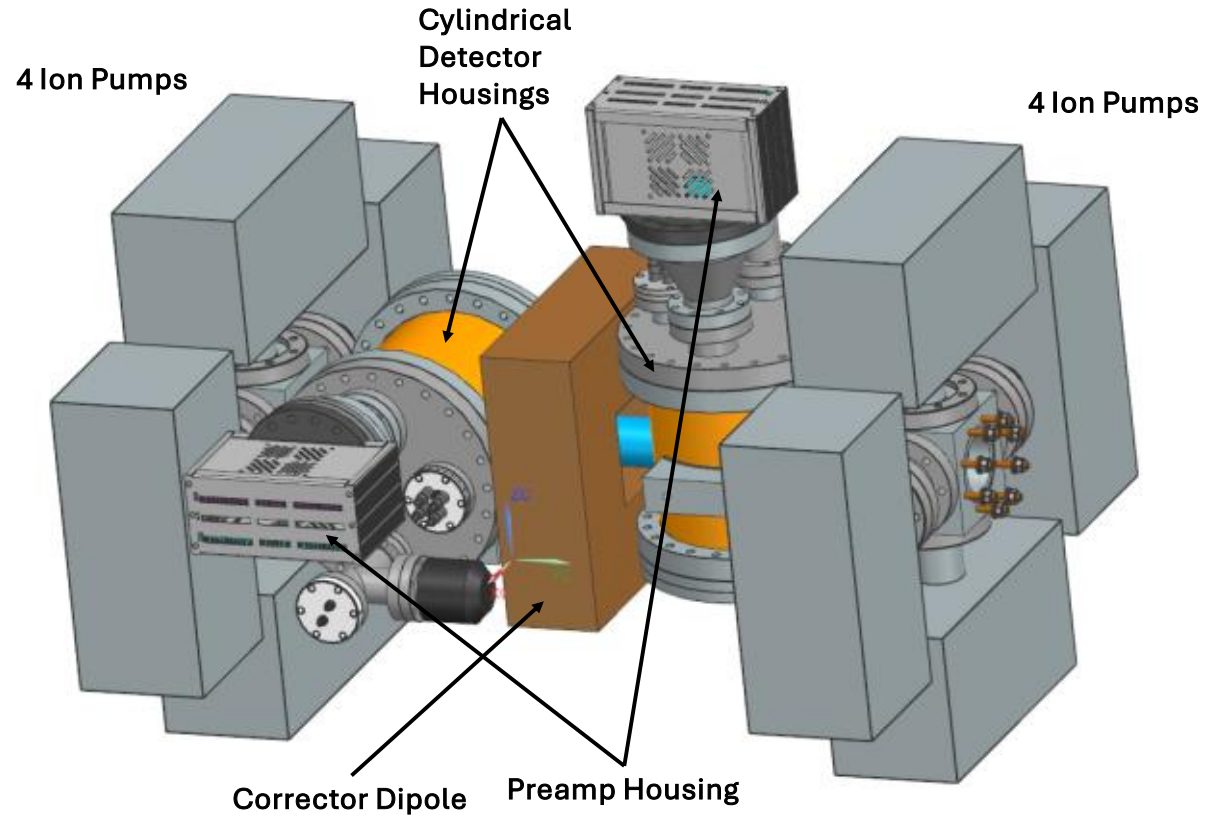

Figure 3: IPM system showing 8 ion pumps, horizontal and vertical detector housings, and the corrector dipole.

Inside each cylinder is an assembly that includes the collection field shaping structure, the wire mesh for shielding the detector from the rf signal of the beam, a pair of microchannel plates (MCPs), and the readout circuit board (Fig. 4). The collection field shaping structure consists of the main plate electrode and three pairs of shaping electrodes to maintain the uniformity of the field. The rf shielding mesh is to prevent beam rf-induced signals and is a Precision eForming MN3 Ni mesh with an 86% open fraction. After the mesh is a pair of Photonis 30444PS Extended Dynamic Range MCPs separated by a 400-µm-thick Cu-coated Kapton spacer. The MCPs are 97 mm × 79 mm × 1 mm thick with 25-µm diameter pores spaced by 32 µm with an 8° tilt. They each have a resistance of ~2 MΩ and a maximum gain of $10^4$. The final element in the structure is the readout strip board which is an $Al_2O_3$ circuit board with 60 conductive 0.35-mm-wide strips having a center-to-center spacing of 0.5 mm.

To help correct for the orbit distortion caused by the collection field (20 mm horizontal, 11 mm vertical), a dipole corrector is placed between the two cylindrical detector housings. Figure 5 (reproduced from [6]) shows the design of the magnet.

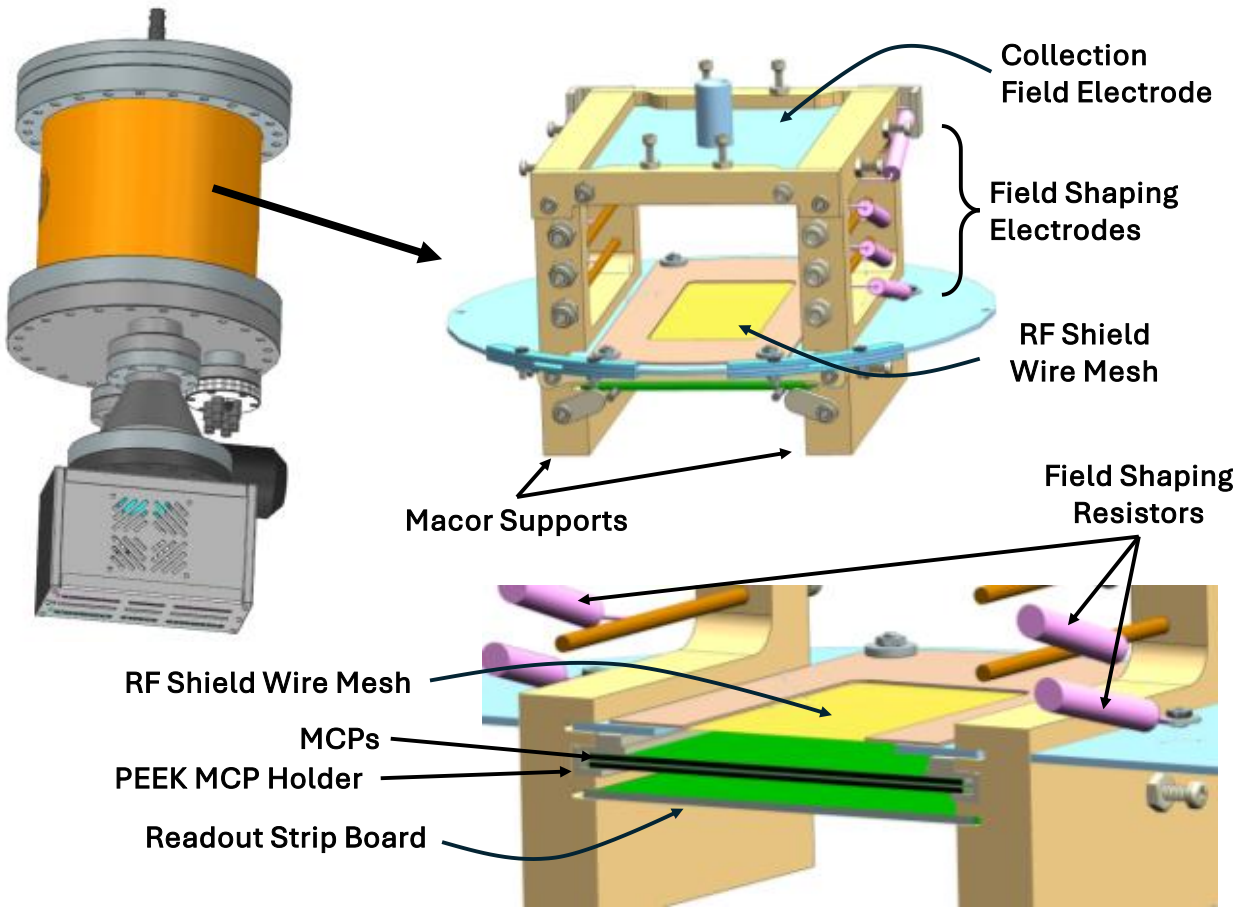

Figure 4: The ion collection system (top right) is located inside the cylindrical housing (top left). In addition to the MCP holder, the various screws are also made from PEEK to avoid arcing.

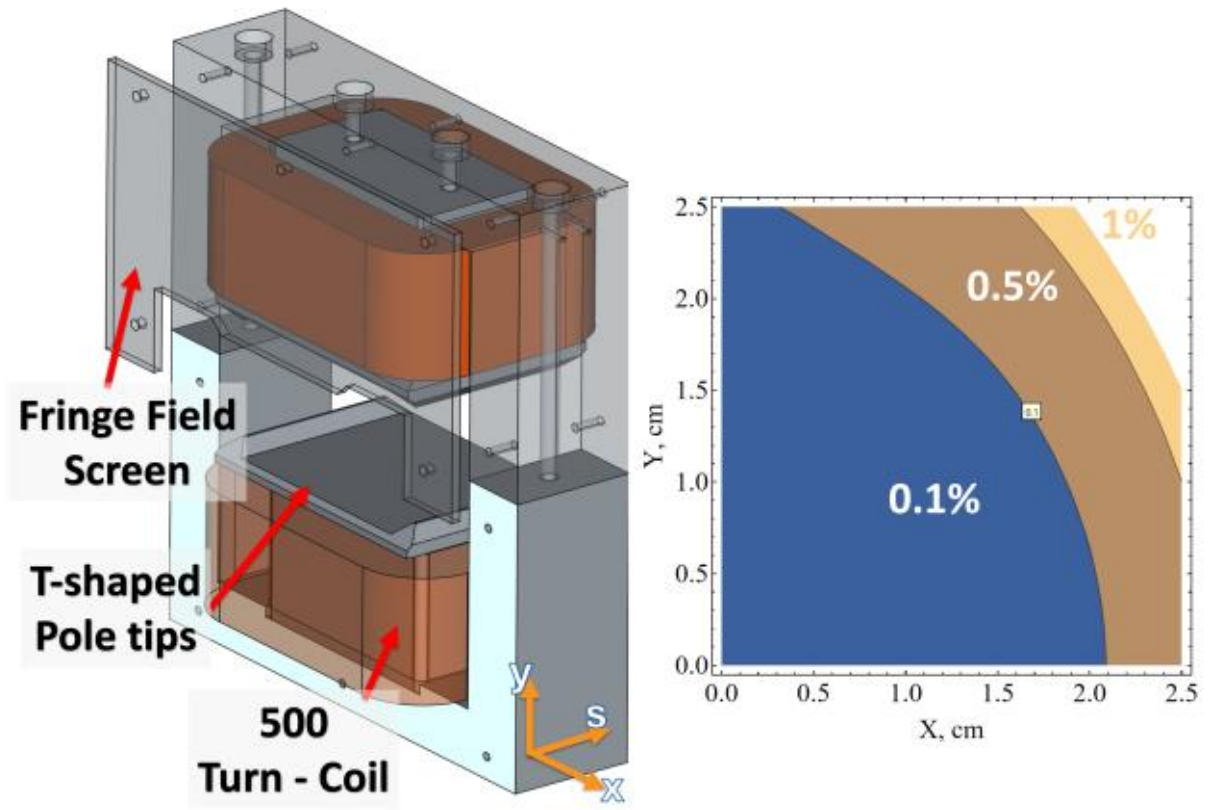

Figure 5: Corrector dipole magnet design for the IPM system and corresponding deviation of the integral field from that at the center. The T-shaped poles improve uniformity. (Figure reproduced from [6])

Without this extra magnet, the orbit deflections utilizing existing corrector magnets would be on the order of 3.2 mm horizontal and 1.8 mm vertical and push the magnets near their field limits. The corresponding values with this extra magnet will only be ~0.6 mm. The magnet will be nominally positioned with a rotation around the beampipe at 45° from horizontal but can be rotated within a range of ~100° to optimize the correction.

## *Gas Injection*

The extremely low design pressure of the IOTA ring results in an extremely small signal from the residual gas ionizations. To mitigate this, a small amount of gas will be injected into the IPM chambers to increase the signal (Fig. 6).

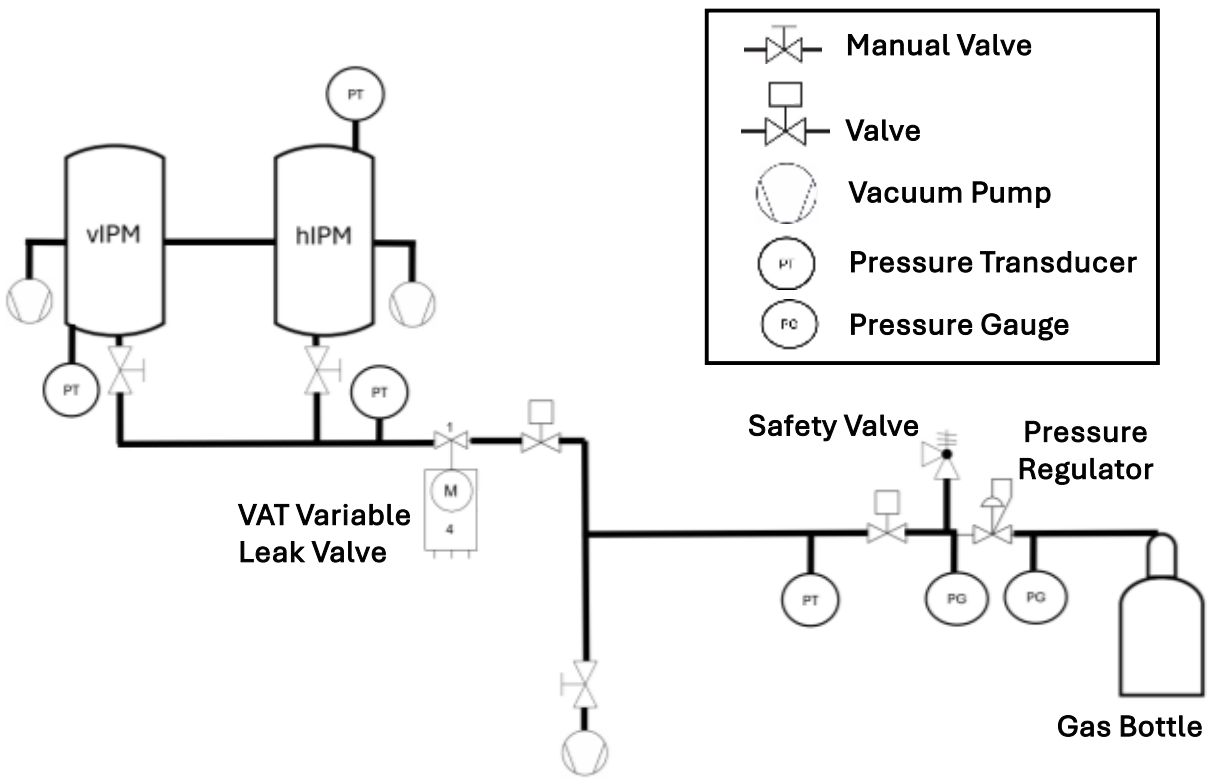

Figure 6: Gas injection diagram.

To maintain the vacuum pressure in the rest of the ring, gas flow impedances must be installed in the beamline on either side of the IPMs (Fig. 7). These impedances are in the form of aperture-restricting tubes known as collimators, that reduce the diameter of the beam pipe from 47.6 mm to 36.5 mm.

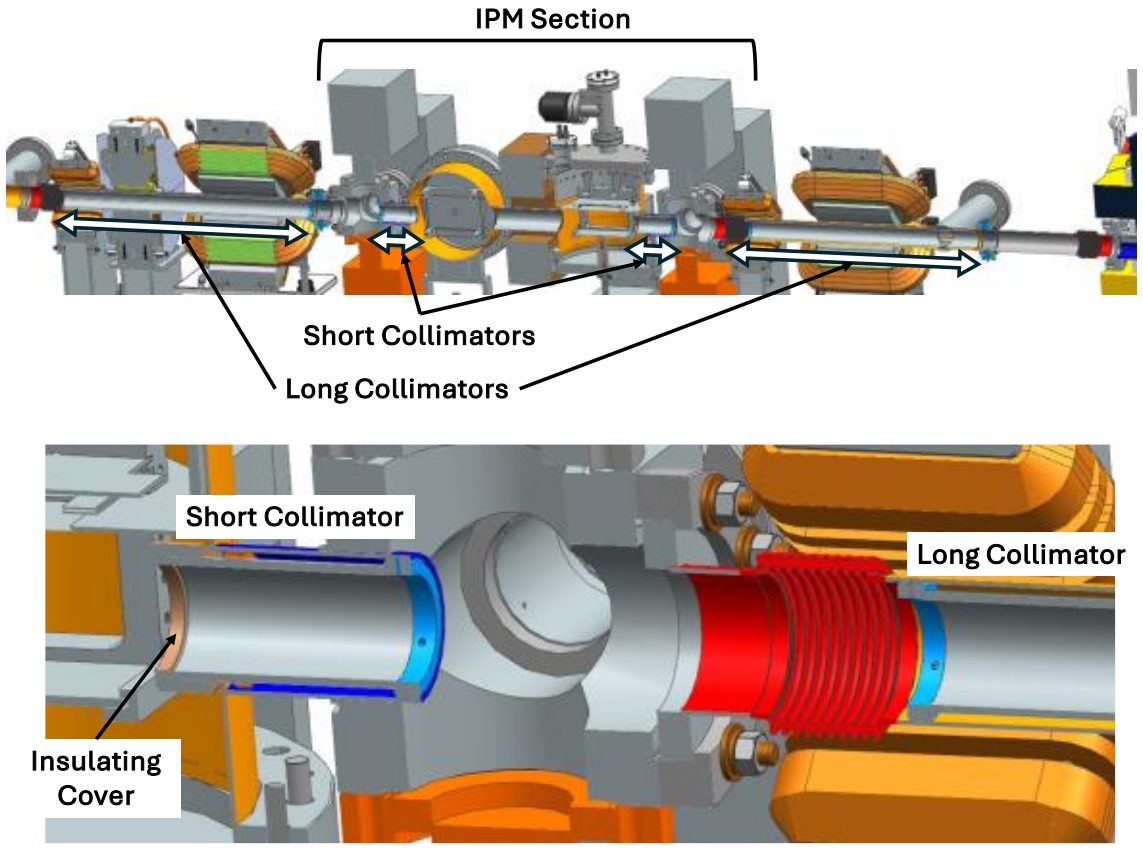

Figure 7: Top) IPM region of IOTA showing the long and short gas collimators to decrease gas leakage out of the IPM detector housings. The short collimators extend into the IPM detector area and thus require an insulating cover to decrease the risk of electrical arcs.

The impact of the gas leak on the proton beam emittance and lifetime, assuming no other effects, was calculated in [6] using an analytical model following [7]. Figure 8 (reproduced from [6]) shows the emittance growth increase and lifetime decrease as the amount of gas (expressed as ions generated per turn) increases. The expected number of ions per turn for a pressure in the IPM region of $10^{-9}$ Torr is ~1800. This situation features two competing needs: more signal, and less degradation of the beam. Ultimately, it will likely need to be adjusted based on the requirements of the particular measurement being made.

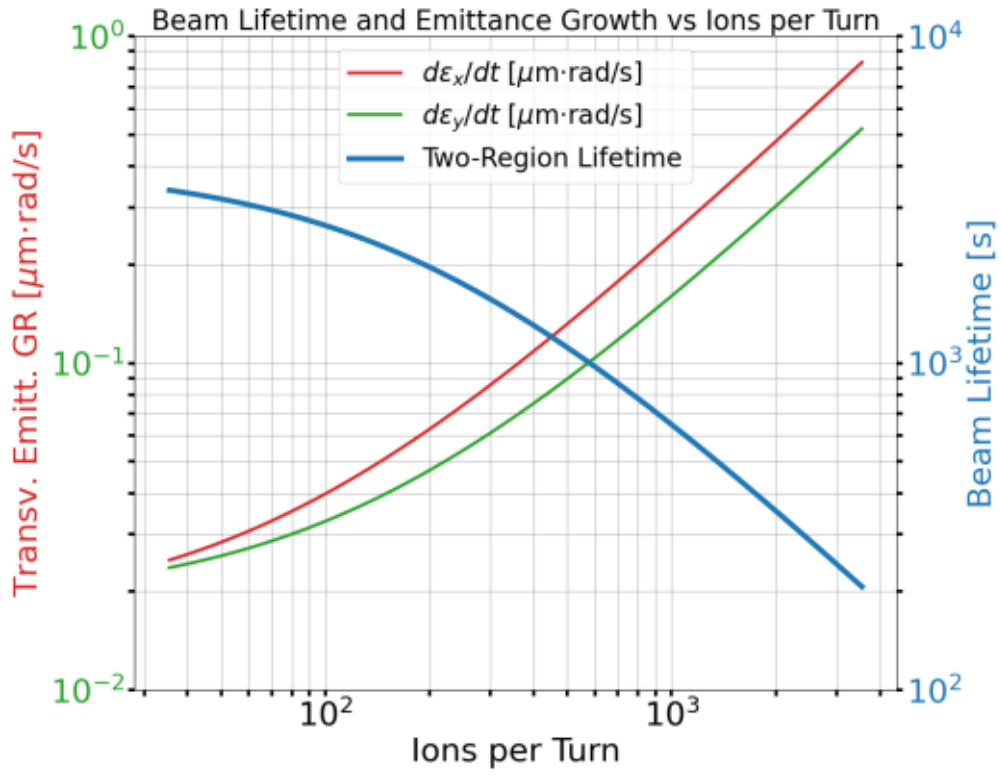

Figure 8: Plot of transverse emittance growth and beam lifetime decrease as the injected gas (ions per turn) is increased. (Figure reproduced from [6])

## *Electrical Design*

The electrical design requires a high voltage (HV) source for the ion collection field, and separate high voltages for each side of each MCP. For the ion collection field, Matsusada AU-30R1-LC HV supplies will be used. Control of these will be through an AutomationDirect Productivity2000 programmable logic controller (PLC).

The voltages related to the MCPs must be in series and will be accomplished via a multi-step voltage divider (Fig. 9). The voltage divider design is a series of 100-V Zener diodes with connection points between each diode and allows the selection of different voltages for each MCP. One could also use multiple stacked HV supplies, but they would need to be multi-quadrant supplies. The choice of a manual divider allows the use of already-owned Bertan 225-03R HV supplies.

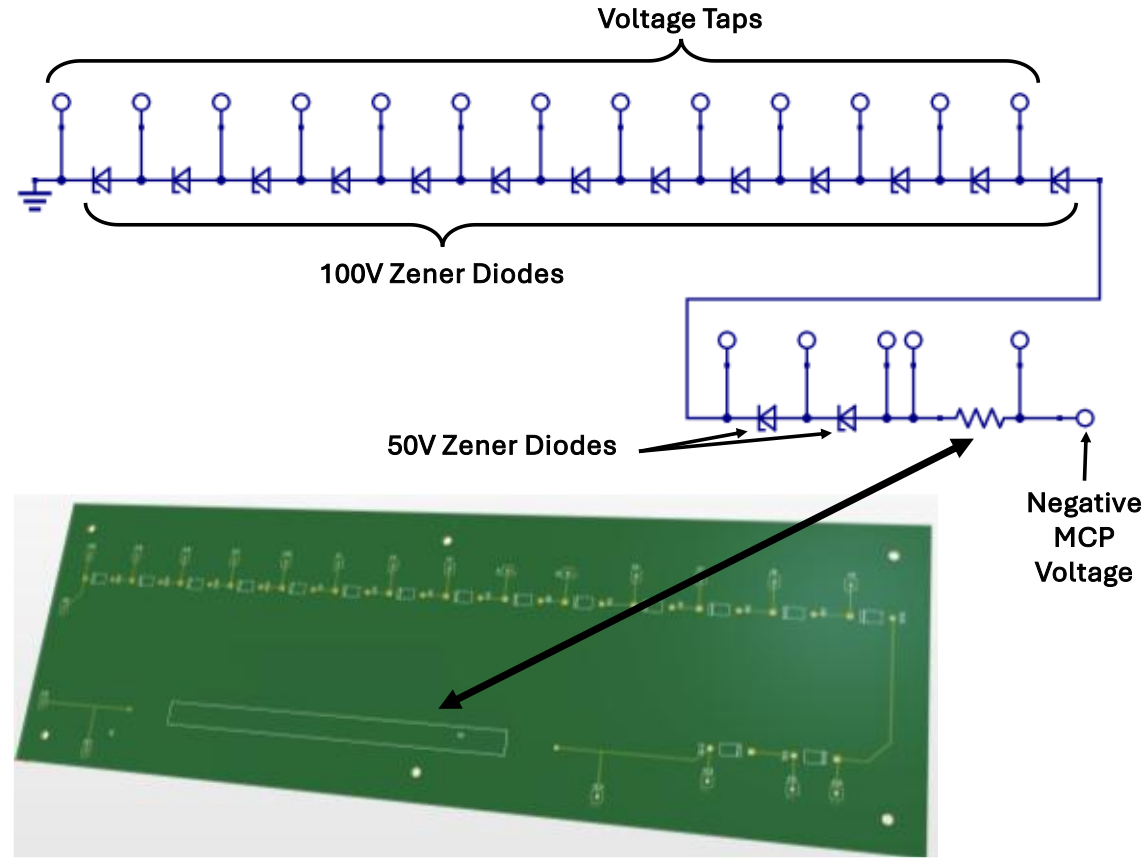

Figure 9: MCP voltage divider circuit. The string of 100-V Zener diodes allows some degree of customizing of the voltage on both sides of both MCPs.

### *Data Acquisition*

The data acquisition system utilizes a current-to-voltage amplifier mounted directly on the vacuum flange to reduce cable capacitance and increase the bandwidth of the amplifier (Fig. 10). The first op-amp section allows an offset to be applied to the signal. This offset is utilized to make use of the full range of the bipolar ADC downstream. Each board has 20 channels.

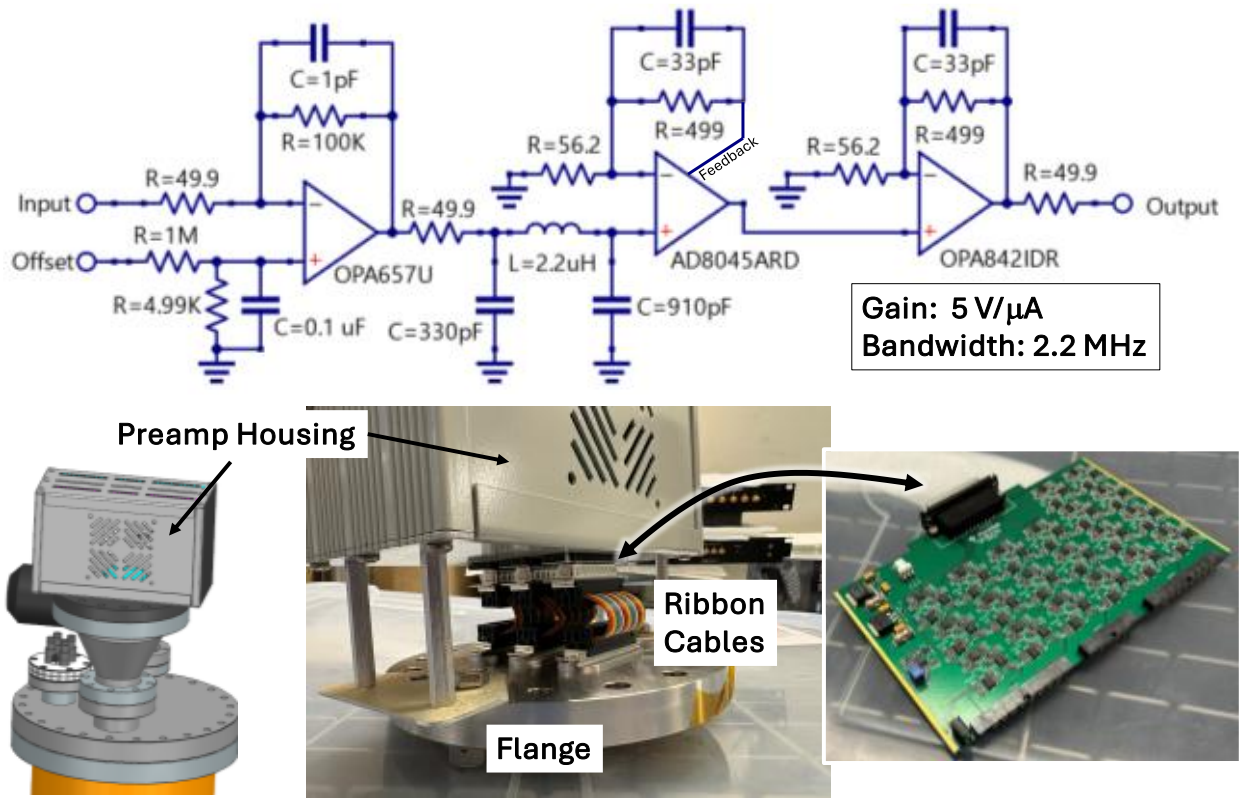

Figure 10: The pre-amp circuit for a single channel consists of three consecutive amplifiers. The boards are mounted on the cylindrical detector housing flange with a short ribbon cable connection.

Signals from the pre-amps are sent out of the enclosure to digitizer boards (Fig. 11) which have 16-channels and digitize at 80 MHz. They are powered via a VME crate, but do not use the VME backplane for data transfer. Instead, data is extracted from the digitizer boards through an ethernet switch and aggregated in a server [8]. This server contains the programs for communication with various instrumentation systems, as well as a Redis database and an interface to the accelerator controls system. The digitizer boards also require timing and trigger inputs to control when data acquisition happens.

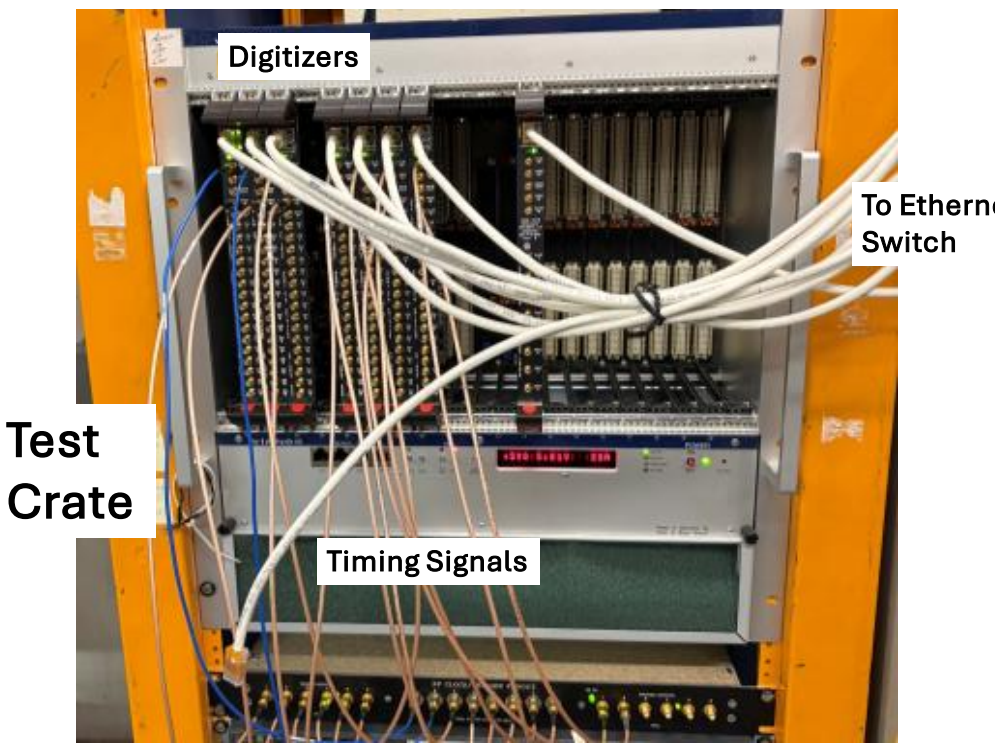

Figure 11: Test crate for evaluating the digitizer firmware and the connection to the server. Timing signals in the test come from a custom timing fanout box.

The digitizer firmware has been designed to enable a flexible acquisition environment (Fig. 12). One can specify either a hardware or software trigger, followed by a user-specified delay. Then a sequence of acquisitions occurs, each of which is an average over a specified number of turns followed by a gap of a specified number of turns. The total number in the sequence is also specified.

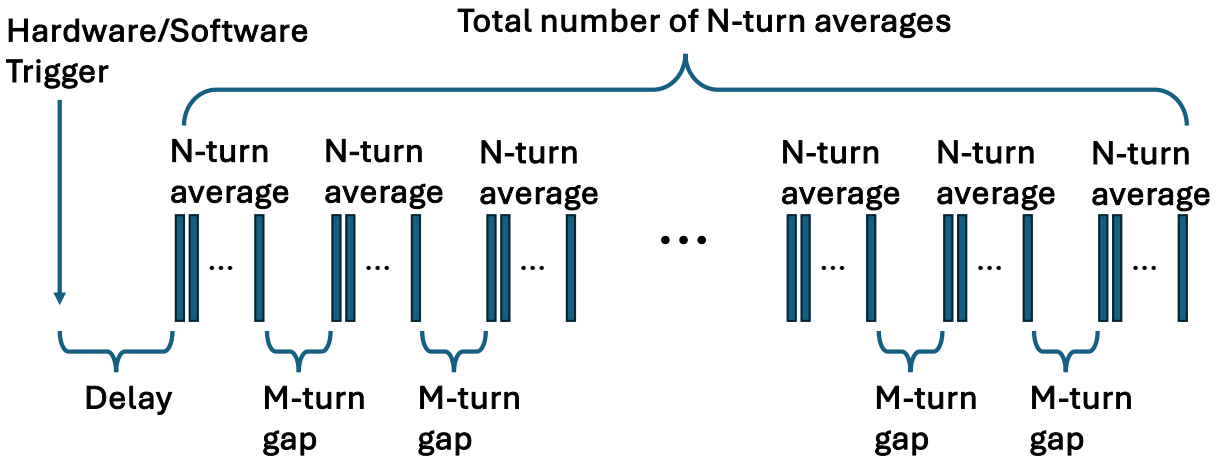

Figure 12: Acquisition sequence showing the adjustable parameters for controlling the data collection.

A high-level application is being designed presently to control the acquisition sequence, as well as controlling the high voltages, and displaying the results.

## SUMMARY

Progress on the IPM detector continues across several fronts. Some vacuum components have been completed, and active in-house machining of the detector internals is currently underway. The ion pumps and controllers have been procured. On the electronics side, the MCPs and HV supplies are already in-house, digitizers are complete, and preamps are partially finished. Voltage dividers are partially constructed, and cables are in house as well. Work on the gas injection components is ongoing, and the trim dipole components are on hand with only some 3-D printing and the winding of the coils remaining. Development of the control application has also begun.

Overall, the IPM project remains on track for a target installation in IOTA completion date of late 2026 to early 2027.